\documentclass[a4paper,5pt]{article}
\usepackage[maxbibnames=99]{biblatex}
\usepackage[text={170mm,240mm},centering]{geometry}
 
\usepackage{graphics,graphicx}
\usepackage{subfig}
\usepackage{float}
\usepackage{booktabs}
\usepackage{longtable}
\usepackage{supertabular}
\usepackage{tabularx}
\usepackage{multirow}
\usepackage{array}
\usepackage{tabu}
\usepackage{mathtools}
\usepackage{hyperref}
\usepackage{accents} 

\usepackage[ruled,vlined]{algorithm2e}

\usepackage{amsmath}
\usepackage{amsfonts}
\usepackage{amssymb}
\usepackage{siunitx} 

\usepackage{enumerate}
\usepackage{sectsty}
\usepackage[affil-it,auth-sc]{authblk}
\usepackage{color}
\usepackage[dvipsnames]{xcolor}

\sectionfont{\normalsize}
\subsectionfont{\normalsize}

\usepackage{tikz}
\usetikzlibrary{arrows,decorations,backgrounds,shapes, snakes}
\usetikzlibrary{positioning}
\usetikzlibrary{matrix} 
\usepackage{varwidth}
\usepackage[most]{tcolorbox}
\usetikzlibrary{shapes.geometric,arrows.meta,decorations.markings}

\usepackage{fancyhdr}            
\fancypagestyle{plain}{%
  \fancyhead{}                                     
  \fancyfoot[CE,CO]{\thepage}       
  \fancyhead[CE,CO]{\textcolor[rgb]{0.64,0.15,0.15}{Published in  \emph{International Journal of Solids and Structures} (2026), \textbf{341}: 114332.\\
  doi: https://doi.org/10.1016/j.ijsolstr.2026.114332}}
\setlength{\headheight}{14.5pt}}

\begin{document}

\renewcommand\Affilfont{\itshape}
\setlength{\affilsep}{1em}
\renewcommand\Authsep{, }
\renewcommand\Authand{ and }
\renewcommand\Authands{ and }
\setcounter{Maxaffil}{2}

\title{
Parametric excitation of rotational soft modes of \\ the buckled discrete elastic ring
}
\author[1]{P. Koutsogiannakis\footnote{Corresponding author: panagiotis.koutsogiannakis@colorado.edu}}
\author[1]{M. Ruzzene}
\affil[1]{P. M. Rady Department of Mechanical Engineering, University of Colorado Boulder, 1111 Engineering Dr, Boulder 80310, CO, USA}

\maketitle

\begin{abstract}
In the present study, we explore the existence and characteristics of a rotating vibrational mode in a discrete buckled ring that manifests under transverse harmonic forcing. The study is conducted through mathematical analysis complemented by numerical simulations. 
The ring is composed of uniformly distributed concentrated masses connected by linear and torsional springs and supported by an elastic foundation.
The buckled state of interest results from homogeneous in-plane compression and is characterized by vibrational modes that have a rotating nature because of the rotational symmetry of the system.
The resulting vibrational mode, called a rotational soft mode, is characterized by large amplitude rotations of the ring deformation while keeping its angular momentum zero. 
It is shown that rotational soft modes can be activated via transverse harmonic forcing through parametric resonance, which is possible only when nonlinearities are introduced to the foundation. In the absence of nonlinear behavior, transverse vibrations of the ring cannot lead to excitation of the rotational mode.
Analysis of a constrained kinematics model shows that linear stiffness of the foundation leads to decoupling of the rotational mode from transverse vibrations. When the foundation is considered piecewise linear, with different stiffness values in compression and in extension, it is shown that three different behaviors are possible for the dynamic response of the forced ring: i) no rotation, ii) a limit cycle within a limited finite rotation angle range, iii) induced rotation between subsequent equilibrium angles.
The present study contributes to the understanding of the activation and control of soft modes in buckled structures and may lead to actuation applications in soft robotics, as well as advanced manipulation strategies of molecular systems in the context of nanomechanics.
\end{abstract}

\vspace{5 mm}

\maketitle

\section{Introduction}\label{sec:intro}

Dynamic systems with constant potential energy in a compact range of their configuration parameters exhibit zero-energy modes. This means that there exists a continuous path in the configuration space of the system for which the system has a finite response to infinitesimal dynamic input that produces no work; see \cite{tarnai2003zero, guest2011zero, Schenk2014}. Zero-energy mechanisms have been proposed for applications in actuation (\cite{kulic2005, baumann2018motorizing, deng2024light}) and manipulation (\cite{mustaza2018stiffness}) in soft robotics, as well as passive vibration isolation (\cite{carrella2007static, LIU20133359, ZHOU2015121, 10.1115/1.4036465}).

In the context of elastic structures, zero-energy modes can be obtained by taking advantage of geometric symmetries.
The simplest example of a dynamic system with such symmetry is a spring pendulum (\cite{kane1968class, bayly1993empirical, lee1999second, de2018energy}). In the absence of gravity, the pendulum can be rotated around the hinge freely, as a small impulse will result in a constant rotation of the mass around the hinge. One characteristic of this example is the axisymmetric nature of its potential energy profile. In more detail, the configuration of the mass-spring system is given by two independent parameters, namely the radial displacement and the rotation of the mass. When the mass is displaced in the radial direction, the potential energy varies parabolically and has a minimum at the natural length of the spring. However, as the mass rotates around the hinge, the potential energy remains constant. This means that the potential energy is minimized on a circle. When gravity is considered, the continuous rotational symmetry is broken, and the spring pendulum exhibits parametric resonance (\cite{warminski2006autoparametric, duka2019elastic}) to external forcing parallel to the arm, with energy continuously exchanged between the radial and rotational modes \cite{de2018energy}. The spring pendulum response can be periodic, chaotic, or characterized by the absence of rotation; see \cite{kane1968class, bayly1993empirical, lee1999second, broucke1973periodic, VANDERWEELE1996245}.

\begin{figure}
    \centering
    \includegraphics[width=140mm]{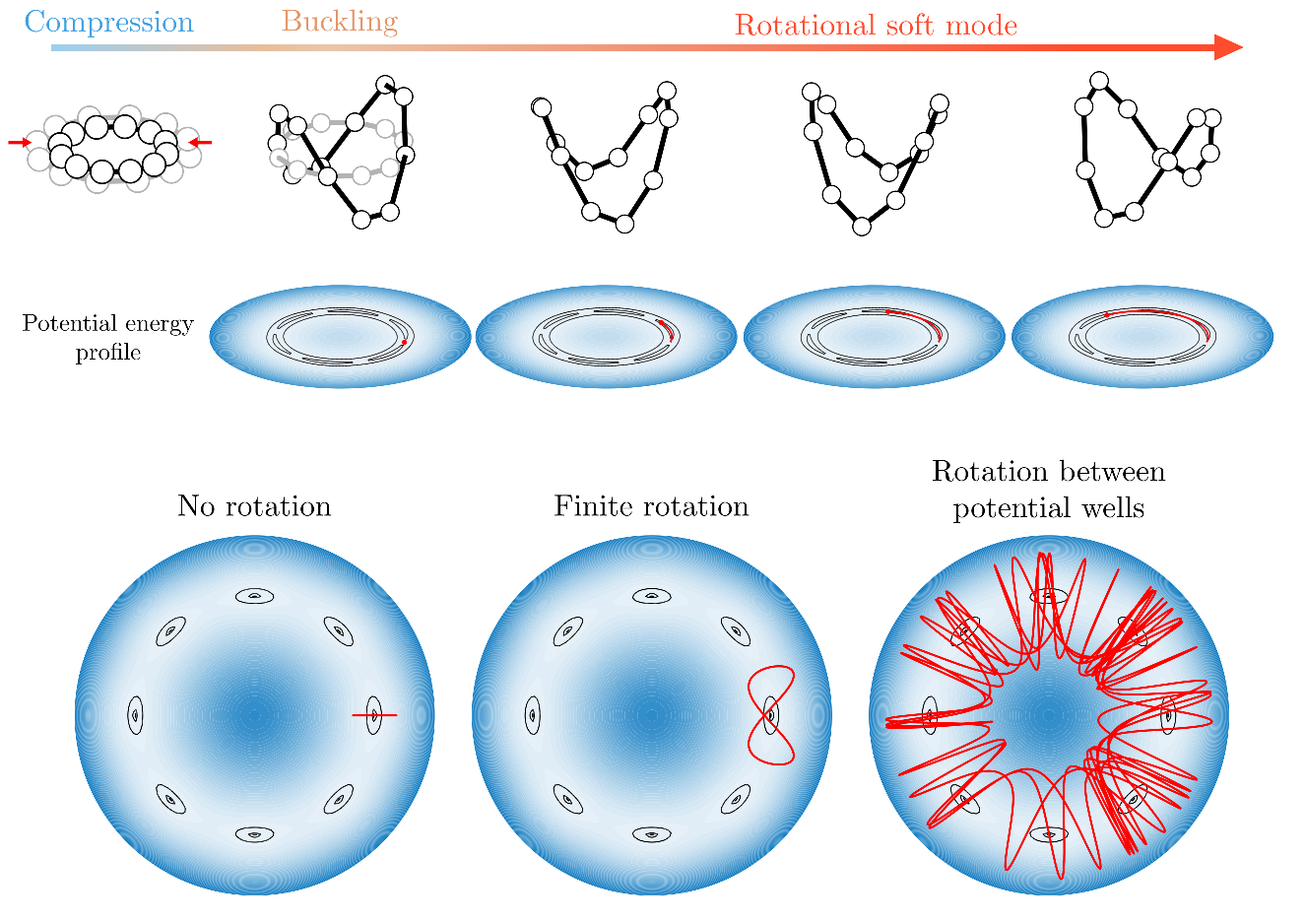}
    \caption{The discrete elastic ring exhibiting a rotational soft mode when compressed beyond the critical point for off-plane buckling. After buckling the potential profile of the ring shows shallow potential wells and the deformation can be rotated with minimal energy input. When the ring is connected to an elastic foundation subject to vertical forcing, three rotation behaviors are possible: no rotation (lower left), oscillating rotation (lower center), or rotation leading to jumps between potential wells (lower right).}
    \label{fig:graphical_abstract}
\end{figure}

Axisymmetric potential energy profiles can also be found in more complex structural systems. For continuous systems, the potential energy for a deformed configuration remains constant as the deformation field rotates around the symmetry axis, leading to zero-energy rotating modes. Structural systems of this kind can be viewed as a simplified, two degree-of-freedom coupled system. In this context, the dynamics resemble that of the spring pendulum, with the primary parameters of the response being the amplitude and rotation of the deformation. When the potential energy symmetry is broken, the rotation mode can be excited indirectly, as, for example, in the case of a bistable plastically deformed disk described in \cite{chibbaro2022chaotic}.

The concept of a rotational soft mode can be illustrated by considering the simple case of a mass connected to a central linear spring in the absence of gravity. Describing its configuration by the radial coordinate $r$ and the rotation angle $\theta$, the potential energy is
\begin{equation*}
\mathcal{V}(r,\theta)=\frac{1}{2}k(r-\ell_0)^2,
\end{equation*}
and is therefore independent of $\theta$. At the equilibrium radius $r=\ell_0$, all values of $\theta$ have the same potential energy, so that the system can move along the rotational coordinate without an associated change in potential energy. This defines an ideal rotational zero-energy mode. When the rotational symmetry is weakly broken, the potential energy develops a small variation with $\theta$, and a finite but small amount of energy is required to move between neighboring orientations. We refer to this situation as a \textit{rotational soft mode}. The buckled discrete ring considered in this work exhibits precisely this behavior: its discrete rotational symmetry produces shallow potential-energy wells associated with different orientations of the buckled deformation.

Stable non-trivial deformation fields can be obtained by buckling instabilities. For example, in \cite{guest2011zero}, a disk made of two fused metal sheets with different thermal expansion coefficients subjected to heating buckles to a stable off-plane configuration. 
Similarly, off-plane deformation of structures can occur with compressive loads. Buckling conditions due to internal compressive loads, as is the case of shrinking, have been studied for continuous annular structures; see \cite{MORIMOTO2020105610, 10.1115/1.4011755}. When compression exceeds the critical value for instabilities to occur, the equilibrium configuration has an off-plane deformation component with finite amplitude and the post-buckling deformation can freely rotate around the symmetry axis. It should be noted that, in contrast to the spring pendulum, this behavior allows the existence of rotation modes with zero angular momentum, since the zero-energy rotation only involves the deformation field, and no mass is transferred with the rotation.

In the present article, the dynamics of a buckled discrete elastic ring is studied. The discrete ring is composed of a finite number of masses arranged in a circle and connected to their nearest neighbors with linear springs, and additional stiffness in out-of-plane deformation is given by torsional springs. The ring is uniformly compressed by forces on the plane of the ring that act on each mass. At the onset of compressive buckling, the system deforms in the off-plane direction, and the discrete nature of the potential energy symmetry leads to the generation of potential wells. As a result, the continuous axial symmetry is broken, implying a finite number of static equilibria, and the buckled configurations are characterized by a low-energy rotational mode. This means that the configuration deformation can be rotated with a small energy input, leading to a rotational soft mode.

In the following sections of the present article, it is shown that rotational soft modes can be activated by placing the discrete ring on an elastic foundation subject to harmonic excitation. The broken symmetry of the potential energy creates a mechanism for energy to be pumped into the deformation rotation, and the combination of parametric resonance and local potential wells implies three possible behaviors:
\begin{itemize}
\item No rotation - when the rotational mode is perturbed from equilibrium, any rotation is quickly attenuated.
\item Oscillation within a local potential well - the rotational mode is confined within the boundaries of a potential well.
\item Oscillation within multiple or all the potential wells - the rotational mode jumps between adjacent potential wells. 
\end{itemize}
When the forcing of the system is enough for the jump to occur, the highly non-linear dynamics of the system results in non periodic behavior. In summary, the concept of exciting the rotational soft mode of the discrete ring through compression, buckling, and subsequent forcing is illustrated in Fig. \ref{fig:graphical_abstract}. The compression procedure that yields the stable non-trivial finite deformation of the ring is shown in the top part, while the three different behaviors of the rotational response are shown in the lower part.

The present study opens up new paths towards control of rotational soft modes. The proposed formulation describes the response of the elastic ring in terms of nondimensional parameters, allowing the results to be interpreted across system with different characteristic sizes and mechanical properties. This provides flexibility in selecting the operating scale of a physical implementation of the system while retaining the underlying mechanism for exciting rotational soft modes. On the other hand, the behavior maps presented in the present articles can be used to find the frequency and amplitude of forcing that will result in a certain behavior. Such scalability may be useful in the design of soft actuators with advanced motion control (see for example \cite{10.1098/rspa.2017.0364}), and at smaller scales, reconfigurable systems for nanomechanical or molecular manipulation (\cite{bedroud2015axisymmetric, 10.1007/s00542-016-3210-y}).


The present article is organized as follows: First, the governing equations for the dynamics of the discrete elastic ring are derived in Section \ref{sec:annulus_dyn}. Next, the conditions that lead to off-plane buckling of the system and the rotational soft mode are studied in Section \ref{sec:buckle}. Lastly, the three different dynamic behaviors of the buckled ring are discussed in Section \ref{sec:resonance}. To assist the readability of the following passages, a nomenclature section is provided in Appendix \ref{app:nomenclature}. The nomenclature is split in two tables, Tables \ref{tab:dimensional} and \ref{tab:nondimensional}, separating the dimensional from the nondimensional quantities and definitions.

\section{Dynamics of the elastic ring} \label{sec:annulus_dyn}

\begin{figure}
    \centering
    \includegraphics[width=0.8\linewidth]{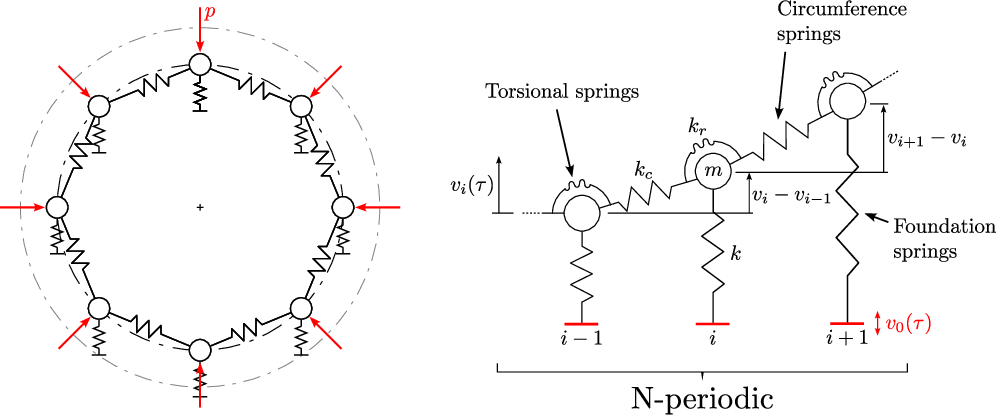}
    \caption{Schematic of the studied system. $N$ masses are uniformly arranged in a circle and connected with their nearest neighbors with linear springs of stiffness $k_c$. At every mass there is a torsional spring of stiffness $k_r$ that acts as the masses are displaced with respect to each other. All the masses are connected to a common ground via linear springs of stiffness $k$. When the ratio of the actual length vs the natural length of the circumferential linear springs exceeds a critical value, the system buckles and deforms in the off-plane direction. In the buckled configuration, the system exhibits a zero-energy rotational mode.}
    \label{fig:ndof_schematic}
\end{figure}

The discrete elastic ring is made up of a network of $N$ point masses $m$, uniformly arranged in a circle on the horizontal plane. The masses are constrained to only have vertical displacements $u_i$. The point masses are connected to their nearest neighbors via linear springs of stiffness $k_c$, and natural length $\ell_0$ and each individual mass is connected to a common foundation via springs of stiffness $k$, as shown in Fig. \ref{fig:ndof_schematic}. The circumferential linear springs are connected via torsional springs $k_r$ that act as the masses are displaced off-plane (Fig. \ref{fig:ndof_schematic} right). If the distance between neighboring masses is $\ell=\ell_0 (1+p)$, then the compression ratio of the ring can be defined as
\begin{equation}
    p=\ell/\ell_0 - 1.
\end{equation}
It is noted that $p$ assumes values in the range $(-1,0]$, with $p=0$ corresponding to no compression and $p=-1$ to full compression degenerating the system to a singular point.
The configuration of the system is given by the nondimensional vertical displacements $v_i=u_i/\ell_0$. A harmonic vibration  $u_0(t)$ of the foundation is considered and given by
\begin{equation}
    u_0(t) = \hat v_0 \ell_0 \sin(\omega t),
\end{equation}
where $\hat v_0$ is the nondimensional amplitude of the motion, normalized by the natural length of the circumferential springs, $\omega$ is the angular frequency of the motion, and $t$ is the dimensional time.

The kinetic and potential energies of the system are
\begin{equation} \label{energies}
    \mathcal T = \sum_{i=1}^N \frac{m}{2} \ell_0^2 \dot v_i^2, \qquad \mathcal{V} = \sum_{i=1}^{N} \left\{ \frac{k}{2} \ell_0^2 (v_i-v_0)^2 + \frac{k_c}{2} \ell_0^2 \left(\lambda_i-1\right)^2 + \frac{k_r}{2}(\phi_{i+1} - \phi_{i})^2 \right\},
\end{equation}
respectively, where a dot over a quantity denotes differentiation with respect to time $t$, and
\begin{equation}
    \lambda_i = \sqrt{(1+p)^2 + (v_i-v_{i-1})^2}, \qquad \phi_i = \tan^{-1}\!\left(\frac{v_{i}-v_{i-1}}{1+p}\right)
\end{equation}
are the normalized distance between two consecutive masses, and angle which the circumferential springs form with the horizontal line respectively. Further, linear viscous damping is considered, implemented by the non-conservative force $-c\,\ell_0\,\dot{v}_i$, where $c$ is the damping coefficient.

By using the principle of least action and defining the following non-dimensional quantities
\begin{equation}
    \qquad q = \frac{k_r}{k \ell_0^2}, \qquad w = \frac{k_c}{k}, \qquad \rho = \frac{m \omega^2}{k}, \qquad \zeta = \frac{c \omega}{k}, \qquad \tau=\omega t,
\end{equation}
the governing equations of the forced discrete elastic ring can be written in a non-dimensional form
\begin{equation} \label{eq:eom_nd}
\begin{aligned}
    \rho \accentset{\ast\ast}{v}_i + \zeta \accentset{\ast}{v}_i + v_i + w \!\left( 1-\frac{1}{\lambda_i} \right) (v_i-v_{i-1}) & - w\!\left(1-\frac{1}{\lambda_{i+1}}\right) (v_{i+1}-v_i) - q\, (1+p) \left( \frac{1}{\lambda_{i+1}^2} + \frac{1}{\lambda_i^2} \right)(\phi_{i+1}-\phi_i) \\ & + q \frac{1+p}{\lambda_i^2}(\phi_i-\phi_{i-1}) + q \frac{1+p}{\lambda_{i+1}^2}(\phi_{i+2}-\phi_{i+1}) = v_0,\quad i=1,\dots,N.
\end{aligned}
\end{equation}
In the equations above, the notation $\accentset{\ast}{(\,\cdot\,)}$ represents differentiation with respect to the non-dimensional time $\tau$.

The discrete model considered here is intentionally restricted to vertical translational degrees of freedom of the masses, with circumferential extension represented by linear springs and changes in the orientation of neighboring links resisted by angular springs. The latter provide bending resistance and should be distinguished from the torsional deformation of a three-dimensional rod about its local axis. Richer discrete formulations, such as Piola--Hencky-type models, may introduce independent rotational degrees of freedom and account explicitly for torsional and shear deformation; see for example \cite{turco2018discrete, turco2020lagrangian}. Such an enrichment would be expected to modify the critical buckling conditions, post-buckling configurations, and quantitative characteristics of the potential-energy wells and dynamic instability regions. At the same time, provided that the additional constitutive terms preserve the rotational symmetry of the homogeneous ring, the symmetry argument underlying the existence of the rotational soft mode is expected to remain valid. Additional torsional and shear degrees of freedom may also introduce new vibrational modes and coupling mechanisms, whose interaction with the rotational soft mode represents an interesting extension of the present study.

\section{Soft mode through buckling} \label{sec:buckle}

As previously stated, finite deformations that lead to a rotational soft mode can be achieved through buckling. The critical values of $p$ for which the system buckles can be found by solving the eigenvalue problem that arises when the system is considered in the quasi-static regime.
The quasistatic equilibrium equations of the discrete elastic ring are derived by Eqs. \ref{eq:eom_nd} by setting the time derivatives equal to zero. Linearizing the equilibrium equations around the trivial configuration $v_i=0$, and grouping together the powers of $p$, the critical values of $p$ are given as the roots of equation
\begin{equation} \label{eq:eig}
    (\mathbf{B}_0 + \mathbf{B}_1 p + \mathbf{B}_2 p^2)\cdot \boldsymbol{\upsilon} = 0,
\end{equation}
where $\mathbf B_0$, $\mathbf B_1$, and $\mathbf B_2$ are sparse banded matrices with nonzero elements
\begin{equation}
    \begin{aligned}
        &B_{0,ii}=1+6q, &B_{1,ii}=2+2w, &B_{2,ii}=1+2w, \\
        &B_{0,ii-1}=B_{0,ii+1}=-4q,\quad &B_{1,ii-1}=B_{1,ii+1}=B_{2,ii-1}=B_{2,ii+1}=-w,\quad &B_{0,ii-2}=B_{0,ii+2}=q.
    \end{aligned}
\end{equation}

\begin{figure}
    \centering
    \includegraphics[width=130mm]{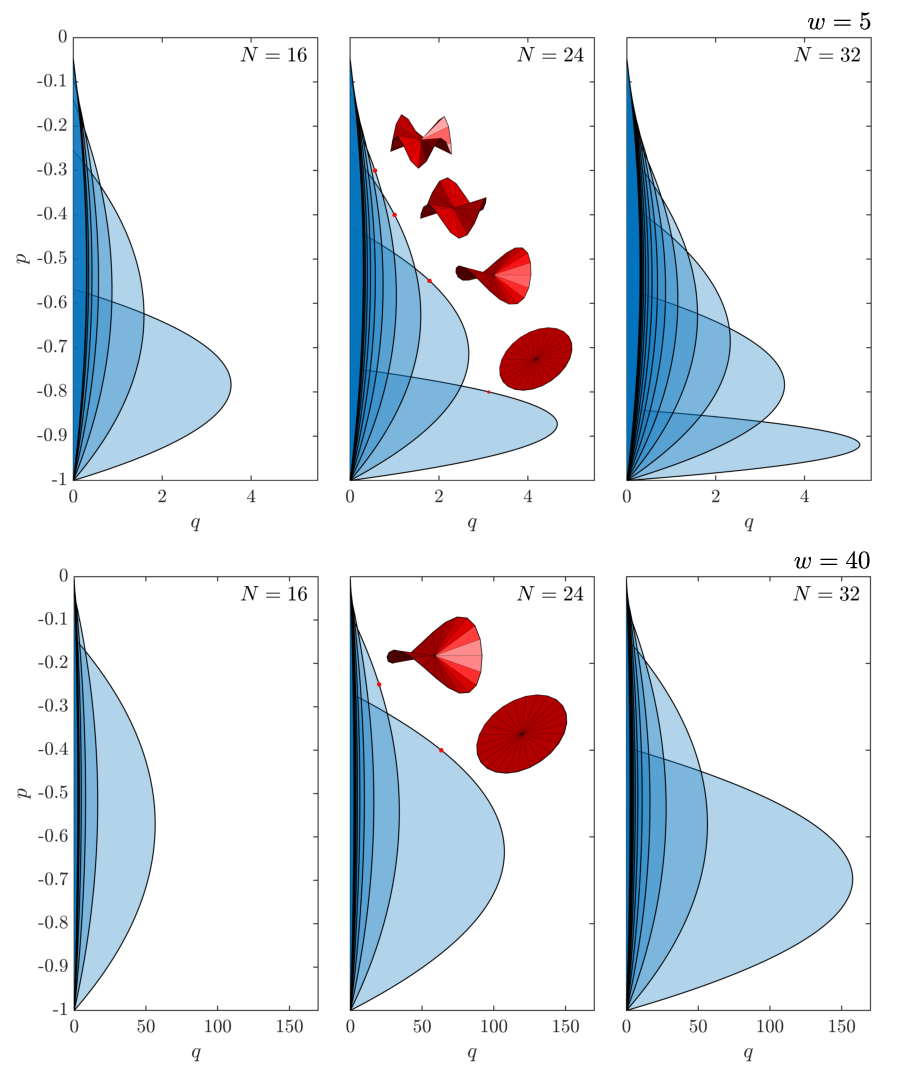}
    \caption{Critical values of the compression parameter $p$ for $N=\{16,24,32\}$ and $w=5$ (top), and $w=40$ (bottom). The trivial configuration becomes unstable when the compression ratio exceeds the first critical value, meaning for all points in the shaded area. As the critical curves intersect, the buckling mode can be selected by choosing the properties of the springs accordingly. For the case $N=24$ some of the buckling modes are shown (red deformed disks).}
    \label{fig:crit_compres}
\end{figure}
The solution of the eigenvalue problem is characterized by three quantities, a rigid body translation $z$, the amplitude of the deformation $A$, and the rotation of the deformation around the center of the ring $\theta$. Then, the eigenmodes are given mathematically in the form
\begin{equation} \label{eq:eig_sol}
    v_i = z + A\, \cos(\alpha\, n\, i + \theta),
\end{equation}
where $\alpha=2\pi/N$ is the angle formed between two consecutive masses and the center of the ring, and $n$ is the buckling mode. It is noted that $\theta$ describes only the unique part of the rotation around the central axis and not the full rotation of the deformed shape, meaning that as $\theta$ goes from $0$ to $2\pi$ the deformation rotates $\theta/n$ times. By substituting the eigenmodes of Eq. \ref{eq:eig_sol} in Eq. \ref{eq:eig},  a simplified quadratic equation is derived for the critical values of $p$
\begin{equation} \label{eq:eig_uniform}
    \left\{ 1+2 w [1-\cos(\alpha n)] \right\}\ p^2 + \left\{ 2 + 2 w [1-\cos(\alpha n)]\right\}\ p + 1 + [6-8\cos(\alpha n)+2\cos(2\alpha n)]\, q = 0.
\end{equation}

The quadratic nature of Eq. \ref{eq:eig_uniform} implies that, as the compression of the system increases (the value of $p$ moves towards $-1$), the trivial configuration of the system ($v_i=0,\ i=1,\dots,N$) becomes unstable at the root of Eq. \ref{eq:eig_uniform} closest to zero and restabilizes at the second root. The critical curves intersect, resulting in different modes of buckling as the ring is compressed. In Fig. \ref{fig:crit_compres} the critical curves are drawn in the $q-p$ plane, for three different values of $N=\{16,24, 32\}$. The top row shows the critical curves for $w=5$, while the lower row for $w=40$. The critical curves for each mode are shown with black lines, while the shaded regions correspond to parameter pairs that lead to instability of the trivial equilibrium. For $N=24$, the buckled shape of the ring is shown as a red shaded shape (with the masses located at the edges of the outer line).

For some values of $w$, there is a secondary restabilization that occurs in a limited range of $q$. Typically, as the ring is compressed the system passes from a stable trivial configuration to unstable and subsequently from unstable to stable as it crosses the critical curve for the second time. However, in the cases for which the secondary restabilization is observed, as the compression continues the configuration becomes unstable again as the critical curve of a lower mode is crossed. Examples of the double restabilization of the discrete ring system are shown in Fig. \ref{fig:crit_compres}(top center, and top right) for the cases $w=5$, $N=\{24,32\}$. The phenomenon of this \textit{double restabilization} has been found to exist for an elastic beam with one end clamped on a slider and the other pinned on a predefined profile \cite{KOUTSOGIANNAKIS2023104745}. To the knowledge of the authors, this phenomenon is of interest from the viewpoint of structural mechanics and has not been further documented in literature, however, its investigation is beyond the scope of the present article. 

In the post-buckling regime, the equilibrium of the system is given by $(z,A,\theta)=(z_{eq},A_{eq},\theta_{eq})$, as well as the shape of the deformation. Due to the non-linear nature of the governing equations, the assumption of a sinusoidal deformation does not hold as the amplitude $A$ becomes large. For this reason, the two parameters $z$, and $A$ are subsequently calculated as
\begin{equation}
    z = \frac{1}{N}\sum_{i=1}^{N} v_i, \qquad A = \frac{\mbox{max}(v_i) - \mbox{min}(v_i)}{2}.
\end{equation}

\begin{figure}
    \centering
    \includegraphics[width=130mm]{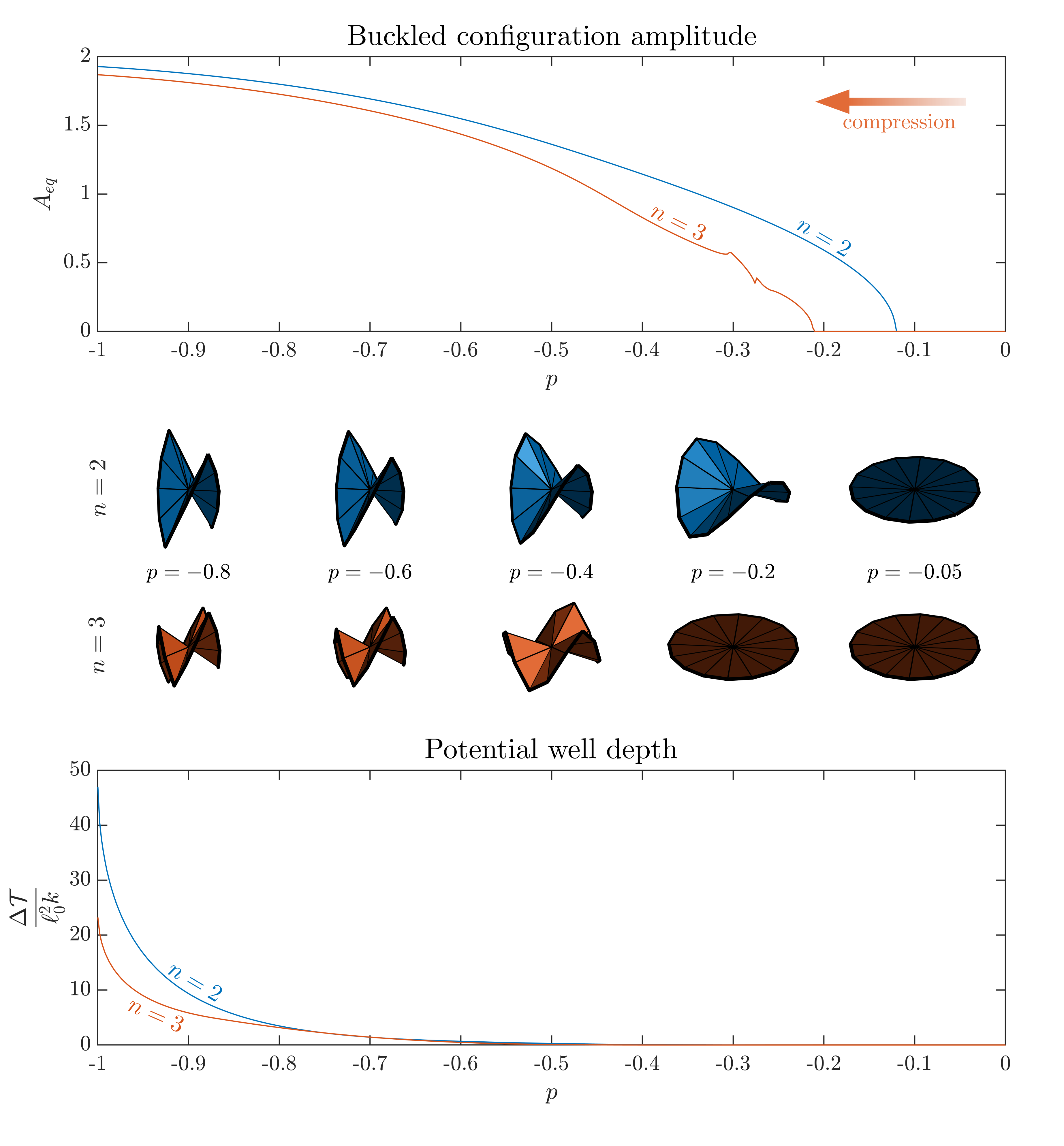}
    \caption{Post-buckling equilibria for $N=16$, $q=5$, and $w=40$ as the ring is compressed ($p\in(-1,0]$). The top and bottom rows show the equilibrium amplitude and the depth of the potential wells formed in the post-buckling regime respectively, for $n=2$ (blue line), and $n=3$ (red line). With compression, both the amplitude $A$ and the potential well depth of the non-trivial equilibrium increase. The potential well increases nearly exponentially while the increase in amplitude diminishes as the value of $p$ tends to $-1$.}
    \label{fig:equil_ampl}
\end{figure}

In Fig. \ref{fig:equil_ampl}(top), the equilibrium amplitude $A_{eq}$ as a function of $p$ is shown for $N=16$, $q=5$, and $w=40$ for the first two modes ($n=2$ shown with blue, and $n=3$ with red). It is observed that for $n=3$ the amplitude does not vary smoothly in the entire range $[-1,p_{cr})$. This is attributed to the interplay between the first and third modes caused by the non-integer $N/n$ ratio. In both cases shown, near the critical compression ratio $p_{cr}$ the amplitude increases dramatically, while at higher compression ratios the amplitude rises more slowly. This is due to the stiffening of the system as the nonlinear terms in Eq. \ref{eq:eom_nd} become significant. Furthermore, select equilibria are shown in Fig. \ref{fig:equil_ampl}(center) for $p=\{-0.05, -0.2, -0.4, -0.6, -0.8\}$, and $n=2$ (shaded with blue) and $n=3$ (shaded with red).

Another important aspect of the quasistatic behavior of the ring system in the post-buckling regime is the fact that the depth of the potential wells rises smoothly with $p$. Near $p_{cr}$ the potential well depth, defined as the difference between the maximum and minimum values of the potential energy as $\theta$ varies in $[0,2\pi]$
\begin{equation}
    \Delta\mathcal T(p) = \mbox{max}(\mathcal T(\theta; p)) - \min(\mathcal T(\theta; p)),
\end{equation}
assumes near-zero values and increases significantly as $p$ gets closer to $-1$. This behavior is shown in Fig. \ref{fig:equil_ampl}(bottom) for the first two modes ($n=2$ with blue, and $n=3$ with red) in the case $N=16$, $q=5$, and $w=40$.

\begin{figure}
    \centering
    \includegraphics[width=130mm]{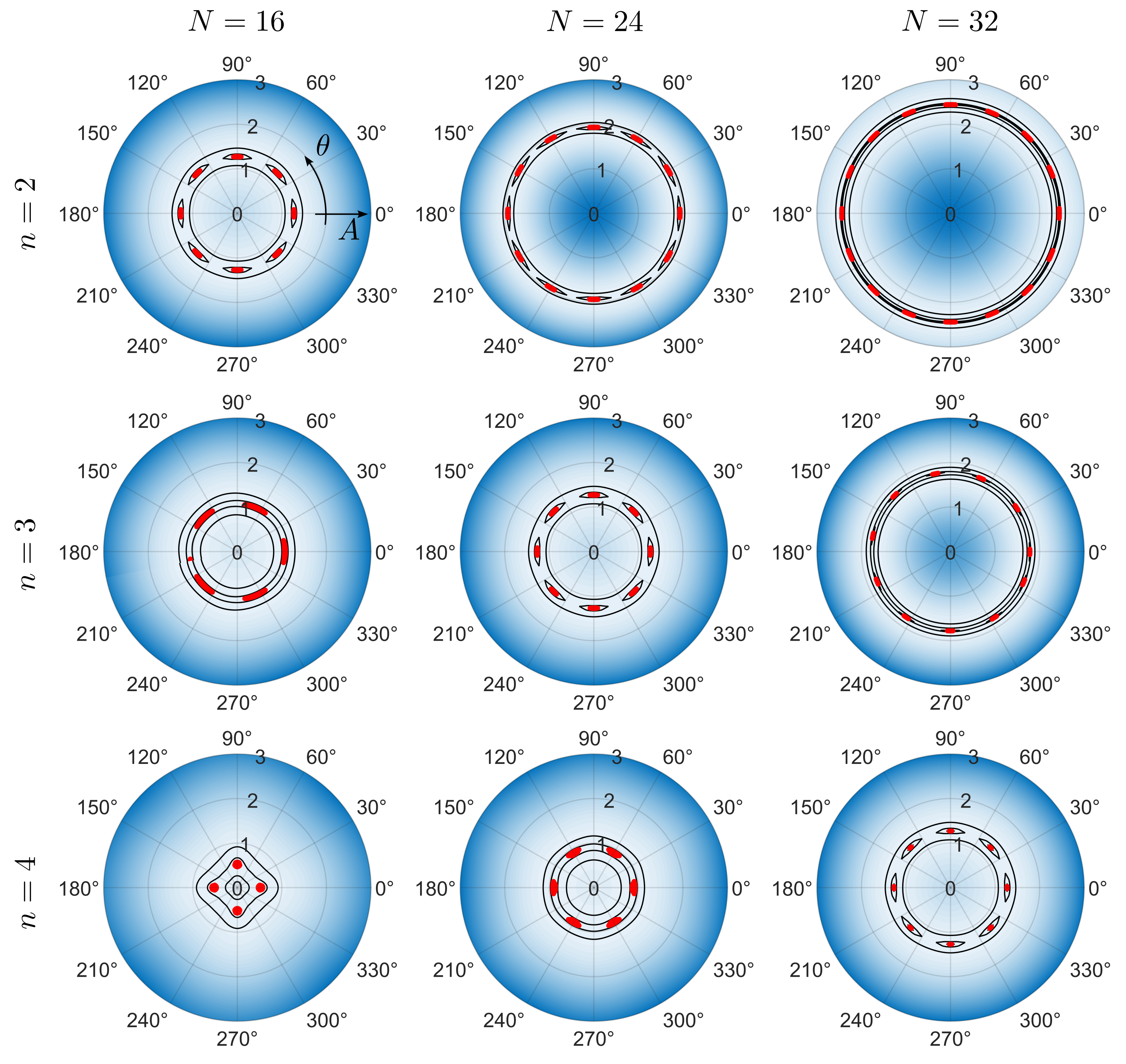}
    \caption{Potential energy of buckled configuration as the deformation is rotated, for $p=-0.5$, $q=5$, and $w=40$. The rows from top to bottom show the potential energy for modes $n=\{2,3,4\}$, and the columns for $N=\{16,24,32\}$. The number of the potential energy minima depends on the buckling mode and the number of masses $N$, shown with red points or lines.}
    \label{fig:potential_energy_minima}
\end{figure}

The variation of the potential energy of the system around the equilibrium configuration can be appreciated by visualizing it on the $A-\theta$ plane. In Fig. \ref{fig:potential_energy_minima} the potential energy of the system is plotted in polar coordinates, with the radial direction corresponding to the amplitude $A$ and the angle to the rotation $\theta$, for $N=\{16,24,32\}$, and $n=\{2,3,4\}$ in the case $p=-0.5$, $q=5$, and $w=40$. In detail, starting from an equilibrium configuration defined as $\theta_{eq}=0$ the deformation is rotated and scaled to span the region $(A,\theta)\in [0,2 A_{eq}]\times[0,2\pi]$. The potential energy minima are shown with red while some contours are shown with thin black lines in order to visualize the shape of the potential energy around the minima.

\section{Parametric excitation of low-energy rotational modes} \label{sec:resonance}

As discussed in Section \ref{sec:buckle}, the behavior of the buckled ring is characterized by the existence of potential wells as the deformation rotates. Similarly to how the rotation of parametric pendulums can be excited by a force acting in the radial direction, the rotation of the ring can be excited by driving the amplitude of the deformation. In order to better understand the mechanism that leads to parametric resonance of the rotational mode a reduced order model is derived. When the amplitude of the deformed shape is sufficiently small, and Eq. \ref{eq:eig_sol} provides a good approximation of the deformation, the response of the system can be given in terms of the time varying parameters $z=z(\tau)$, $A=A(\tau)$, and $\theta=\theta(\tau)$. 

Under this kinematics model given by Eq. \ref{eq:eig_sol}, the equations of motion become
\begin{equation}\label{eq:eom_red_kin}
    \begin{aligned}
        & \rho \accentset{\ast\ast}z(\tau) + z(\tau)-v_0(\tau) = 0, \\
        & \accentset{\ast\ast}A(\tau) + \left(\frac{2-2\cos(\alpha n) +\cos(2 \alpha n)}{\rho}-\accentset{\ast}\theta^2(\tau)\right) A(\tau) + G_2(A(\tau),\theta(\tau)) = 0, \\
        & A(\tau)\accentset{\ast\ast}\theta(\tau)+2\accentset{\ast}A(\tau)\accentset{\ast}\theta(\tau) + 4 \sin^2\frac{\alpha n}{2} \sin(\alpha n) A(\tau) + G_3(A(\tau),\theta(\tau)) = 0,
    \end{aligned}
\end{equation}
where 
$G_2(A(\tau),\theta(\tau))$, $G_3(A(\tau),\theta(\tau))$ are nonlinear functions of $A$ and $\theta$. The derivation of the equation of motion under the constrained kinematics framework is provided in Appendix \ref{app:red_kin_derivation}.
It is noted that the last two equations above resemble the governing equations of the forced spring pendulum \cite{duka2019elastic} and the forced bistable disk \cite{chibbaro2022chaotic}.

An immediate observation is that Eq. \ref{eq:eom_red_kin}a is decoupled from Eqs. \ref{eq:eom_red_kin}b,c, due to the lack of terms involving $z(\tau)$ in the latter two.
A consequence of this is that the vertical forcing of the foundation of the ring cannot excite the rotational mode, as it will only result in a vertical rigid body vibration of the system governed solely by Eq. \ref{eq:eom_red_kin}a. 

In order to facilitate the transfer of energy between the rigid-body component motion of the ring and the rotating mode, the foundation springs should be nonlinear. In this way, a time-varying term, depending on $z(\tau)$, is introduced in Eq. \ref{eq:eom_red_kin}b,c. As a result, the rotation mode can be induced through parametric resonance. 
To allow for this, the springs connecting the ring to the ground are modified by adding a stiffening term that is active when the springs are compressed. More specifically, the foundation springs are considered to have a piecewise linear force-displacement relation and the compression stiffness  $k^{(-)}$ is chosen to be a multiple of the extension stiffness $k^{(+)}$. 
Here, the superscript $(+)$ is used to denote extension of the spring, $v_i-v_0\geq0$, while $(-)$ is used for compression of the spring, $v_i-v_0<0$.
Following this change, the definitions of the non-dimensional quantities $q$, $w$, $\rho$, and $\zeta$ make use of the reference stiffness $k=k^{(+)}$, and the ratio $\gamma$ between the two stiffnesses is defined
\begin{equation}
    \gamma = \frac{k^{(-)}}{k^{(+)}}.
\end{equation}
Then Eqs. \ref{eq:eom_nd} are modified so that the foundation term $v_i-v_0$ is substituted by $\kappa_i (v_i-v_0)$, where
\begin{equation}
    \kappa_i = \begin{cases}
        1, \quad v_i-v_0 \geq0, \\
        \gamma, v_i-v_0<0,
    \end{cases}
\end{equation}
are the normalized stiffness coefficients. The full equations of motion then become
\begin{equation}\label{eq:eom_nd_rev}
\begin{aligned}
    \rho \accentset{\ast\ast}{v}_i + \zeta \accentset{\ast}{v}_i + \kappa_i(v_i-v_0) + w \!\left( 1-\frac{1}{\lambda_i} \right) (v_i-v_{i-1}) & - w\!\left(1-\frac{1}{\lambda_{i+1}}\right) (v_{i+1}-v_i) - q\, (1+p) \left( \frac{1}{\lambda_{i+1}^2} + \frac{1}{\lambda_i^2} \right)(\phi_{i+1}-\phi_i) \\ & + q \frac{1+p}{\lambda_i^2}(\phi_i-\phi_{i-1}) + q \frac{1+p}{\lambda_{i+1}^2}(\phi_{i+2}-\phi_{i+1}) = 0,\quad i=1,\dots,N.
\end{aligned}
\end{equation}

For the results presented here, the compression-to-extension stiffness ratio is selected as $\gamma
=2$. This value is used as a representative stiffness asymmetry sufficient to introduce coupling between the vertical and rotational motions.


In more detail, the mechanism of inducing the rotational mode is the following:
\begin{enumerate}
    \item The vertical forcing of the foundation results in a rigid body motion of the ring.
    \item The non-linear forces of the foundation springs move the location of the trough of the potential energy wells, so that the static equilibrium configuration amplitude $A_{eq}$ depends on the vertical translation of the ring $z$. This means that vertical forcing can induce a change in amplitude $A$.
    \item The coupling of Eq. \ref{eq:eom_red_kin}b and Eq. \ref{eq:eom_red_kin}c creates a feedback loop, implying that a perturbation of $\theta$ around the equilibrium point $\theta_{eq}$ may lead to an amplification of the oscillation of the rotational mode.
\end{enumerate}

The dynamics of the system are demonstrated with a case study. For this purpose the parameters of the system are fixed to the values $N=16$, $p=-0.15$, $q=5$, $w=40$, and the second buckling mode, $n=2$, is considered. To avoid repetition, the ring with this set of parameters is denoted as $\mathcal P$.

\begin{figure}
    \centering
    \includegraphics[width=160mm]{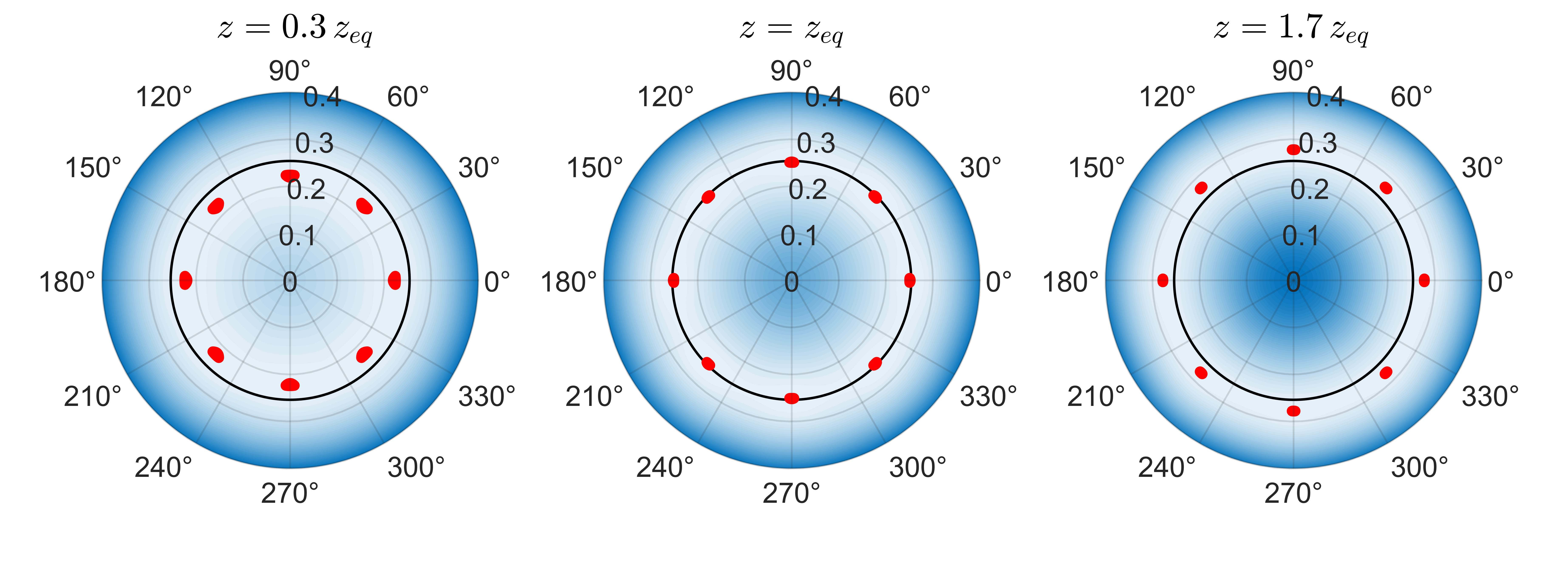}
    \caption{Potential energy on the $\theta-A$ plane for the ring $\mathcal P$ with piece-wise constant stiffness of the foundation springs at different values of $z$. Three cases are shown, $z=0.3\, z_{eq}$ (left), $z=z_{eq}$ (center), and $z=1.7\,z_{eq}$ (right). The potential energy minima are shown with red and the equilibrium amplitude $A_{eq}^0$ corresponding to $z_{eq}$ is shown as a black circle. The amplitude $A$ in each case changes with the value of $z$.}
    \label{fig:potential_var_z}
\end{figure}

The potential energy variation of the ring as $z$ varies is shown in Fig. \ref{fig:potential_var_z} for the set of parameters $\mathcal P$. In particular, the potential energy profile is shown for three values of $z$, $z=0.3\, z_{eq}$ (left), $z=z_{eq}$ (center), and $z=1.7\, z_{eq}$ (right). For all three cases, the potential energy minima are shown in red, while the circle denoting the amplitude of the static equilibrium deformation $A_{eq}^0$ at $z=z_{eq}$ is drawn in black. It is apparent that the deformation amplitude of the potential energy minima is shifting as the value of $z$ changes, providing the mechanism that transforms the vertical vibration of the foundation to an oscillation of the deformation amplitude.

\begin{figure}
    \centering
    \includegraphics[width=150mm]{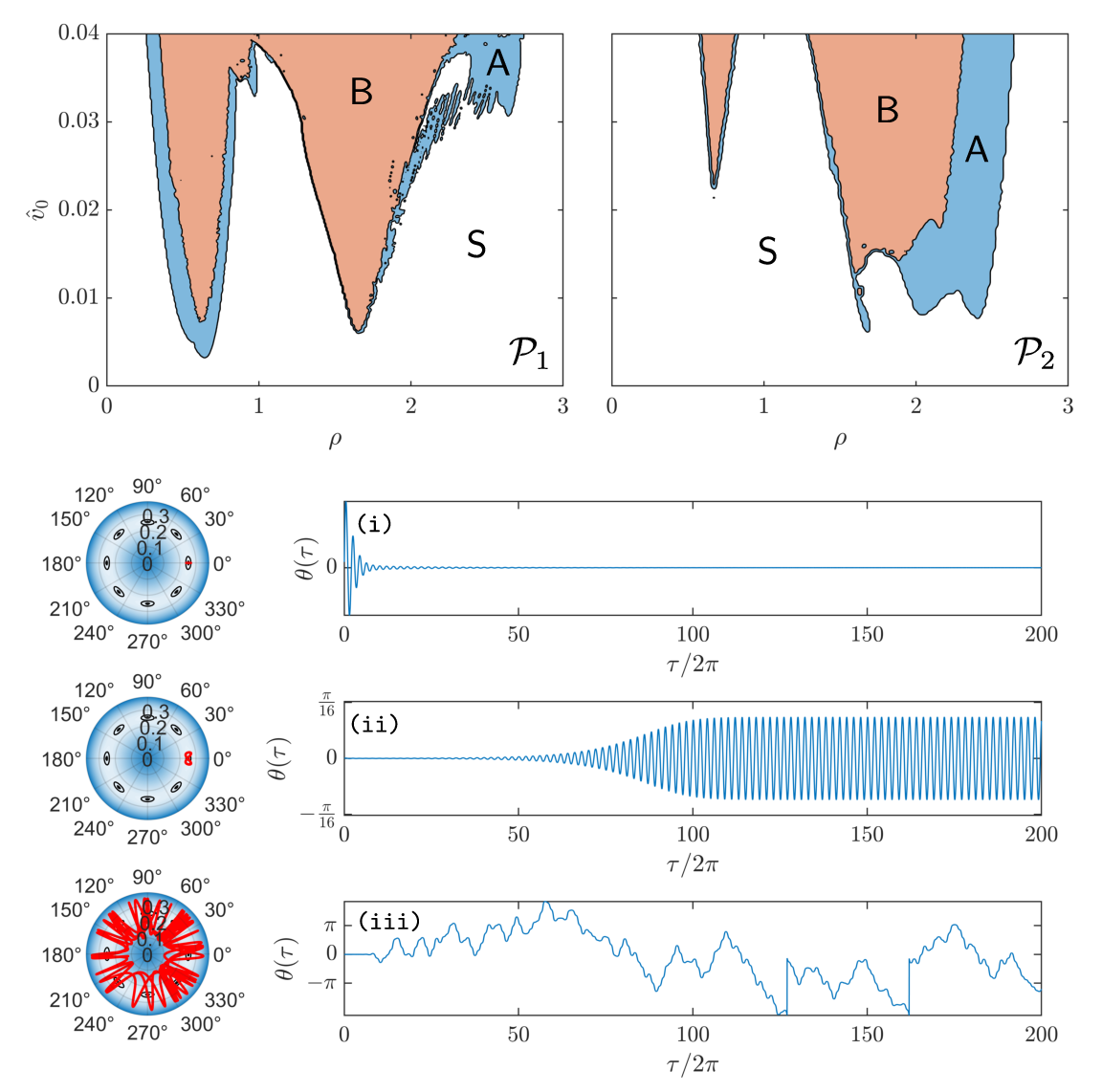}
    \caption{(Top) Regions of the $\rho-\hat{v}_0$ plane where parametric resonance is observed, for the case $N=16$, $n=2$, $p=-0.15$, $q=5$, $w=40$, and $\zeta=0.06$. As the forcing of the system moves into the shaded region, energy is pumped into the soft mode and the deformation of the ring rotates. The region of the $\rho-\hat{v}_0$ plane where the trivial response is unstable is shaded with light blue, and the region where the rotation is large enough for the ring system to jump from one potential well to an adjacent one is shaded with light red. (Bottom) Examples of the three different behaviors observed for $\rho=0.55$. Starting from a small perturbation, the rotational soft mode is attenuated at $\hat v_0=0.0035$ (i), amplified towards a limit cycle at $\hat v_0=0.005$ (ii), jumps between potential wells at $\hat v_0=0.012$ (iii). For these three cases, the response of the ring is overlayed with red over the potential energy profile on the $A-\theta$ plane (bottom left) and the transient time history of $\theta$ is shown for the first 200 periods of the foundation vibration (bottom left). }
    \label{fig:param_resonance}
\end{figure}

For a given set of parameters $\{p, q, w\}$, the resonance of the rotational mode depends on the magnitude $\hat v_0$ and frequency $\omega$ of the forcing. The conditions leading to excitation of the rotational mode can be found in terms of $\hat v_0$ and the non-dimensional frequency $\rho$ by numerical integration of Eqs. \ref{eq:eom_nd_rev} in time; see Appendix \ref{app:numerical} for implementation details. In Fig. \ref{fig:param_resonance}, the effect of forcing on the rotational mode is shown for the ring $\mathcal P$. A small amount of dissipation, $\zeta=0.06$, is considered to eliminate transient effects. It is noted that $\zeta$ depends on the excitation frequency $\omega$ and subsequently on $\rho$. For the purposes of mapping the behavior of the elastic ring we keep $\zeta$ constant, which implies that the dimensional dissipation coefficient $c$ varies. The region of the $\rho-\hat v_0$ plane where the trivial rotation response is unstable is colored, while the region where there is no resonance of the rotational mode is shown in white (region \textsf{S}). 

In the case where the rotational mode is excited, the rotation can be limited within one potential well, exhibiting a limit cycle (region \textsf{A}), or the amplitude of the rotation can exceed the width of the potential wells, leading to jumps between adjacent potential wells (region \textsf{B}). This distinction in the dynamic behavior of the ring is shown in Fig. \ref{fig:param_resonance}(top), where region \textsf{A} is shaded in light blue and region \textsf{B} is shaded in light red. The boundaries between the different regions are not always smooth and around them there exist islands of different behavior. Although this may be true, it is noted that for a fixed value of $\rho$ the response of the system as $\hat v_0$ increases generally passes in order from region \textsf{S}, to region \textsf{A} and finally to region \textsf{B}.
Examples of the three different behaviors of the system are shown in Fig. \ref{fig:param_resonance}(bottom). In particular, for the selected value $\rho=0.55$ the rotational mode can be attenuated when $\hat v_0=0.0035$ (case \texttt{(i)}), stabilize in a limit cycle when $\hat v_0=0.005$ (case \texttt{(ii)}), or have a non-periodic response that jumps between adjacent potential wells when $\hat v_0=0.012$ (case \texttt{(iii)}). 

\section{Conclusions}
The dynamics of the discrete elastic ring on an elastic foundation is studied. When the ring is compressed beyond the critical value for out-of-plane buckling the deformation field can be rotated with minimal energy input, denoting the existence of rotational soft modes. The potential energy as a function of the rotation angle exhibits shallow potential wells. It is shown that when the stiffness of the foundation springs is piecewise linear, with higher stiffness in compression compared to tension, a small harmonic forcing of the foundation can lead to parametric resonance of the rotational mode. A study of the amplitude and frequency of the forcing is performed in order to show the different behaviors of the system. It is found that in terms of the rotational mode the system can have dynamic stability with no rotation or dynamic instability, where the rotational mode is amplified. When the rotational mode is excited, the ring can stabilize on a limit cycle within one potential well, or its amplitude can be large enough that it jumps between adjacent potential wells. 

The formulation of the elastic ring system in a nondimensional manner, allows the description of the system across scales. This facilitates the translation of the observed dynamic regimes to potential physical implementations. By appropriately selecting the characteristic length $\ell_0$, the mass, and the stiffness of the system, the frequency and amplitude of the forcing be selected to achieve a specific dynamic response. Inversely, the maps obtained in Figure \ref{fig:param_resonance} can guide the design of an application once the size and time scale are established. The results therefore support a framework for exploring rotational soft modes in applications ranging from low-energy soft robotic actuation to small-scale reconfigurable and manipulation mechanisms.

\section*{Acknowledgments}
PK and MR gratefully acknowledge funding from the National Science Foundation (NSF) under the grant ”Collaborative Research: Design and Reconfiguration of Curved Surfaces for Targeted Wave Propagation” with award number 2247095.

\section*{References}
\printbibliography[heading=none]

@article{de2018energy,
  title={Energy distribution in intrinsically coupled systems: The spring pendulum paradigm},
  author={De Sousa, MC and Marcus, FA and Caldas, IL and Viana, RL},
  journal={Physica A: Statistical Mechanics and its Applications},
  volume={509},
  pages={1110--1119},
  year={2018},
  publisher={Elsevier}
}

@article{tarnai2003zero,
  title={Zero stiffness elastic structures},
  author={Tarnai, T},
  journal={International Journal of Mechanical Sciences},
  volume={45},
  number={3},
  pages={425--431},
  year={2003},
  publisher={Elsevier}
}

@article{baumann2018motorizing,
  title={Motorizing fibres with geometric zero-energy modes},
  author={Baumann, Arthur and S{\'a}nchez-Ferrer, Antoni and Jacomine, Leandro and Martinoty, Philippe and Le Houerou, Vincent and Ziebert, Falko and Kuli{\'c}, Igor M},
  journal={Nature materials},
  volume={17},
  number={6},
  pages={523--527},
  year={2018},
  publisher={Nature Publishing Group UK London}
}

@article{kane1968class,
  title={On a class of two-degree-of-freedom oscillations},
  author={Kane, TR and Kahn, ME},
  year={1968}
}

@article{bayly1993empirical,
  title={An empirical study of the stability of periodic motion in the forced spring-pendulum},
  author={Bayly, PV and Virgin, LN},
  journal={Proceedings of the Royal Society of London. Series A: Mathematical and Physical Sciences},
  volume={443},
  number={1918},
  pages={391--408},
  year={1993},
  publisher={The Royal Society London}
}

@article{lee1999second,
  title={Second-order approximation for chaotic responses of a harmonically excited spring--pendulum system},
  author={Lee, Won Kyoung and Park, Hae Dong},
  journal={International journal of non-linear mechanics},
  volume={34},
  number={4},
  pages={749--757},
  year={1999},
  publisher={Elsevier}
}

@article{chibbaro2022chaotic,
  title={Chaotic and regular dynamics of a morphing shell with a vanishing-stiffness mode},
  author={Chibbaro, Sergio and Hamouche, Walid and Maurini, Corrado and Vidoli, Stefano and Vincenti, Angela},
  journal={Extreme Mechanics Letters},
  volume={54},
  pages={101755},
  year={2022},
  publisher={Elsevier}
}

@article{10.1115/1.4036465,
    author = {Ishida, Sachiko and Suzuki, Kohki and Shimosaka, Haruo},
    title = {Design and Experimental Analysis of Origami-Inspired Vibration Isolator With Quasi-Zero-Stiffness Characteristic},
    journal = {Journal of Vibration and Acoustics},
    volume = {139},
    number = {5},
    pages = {051004},
    year = {2017},
    month = {06},
    issn = {1048-9002},
    doi = {10.1115/1.4036465},
    url = {https://doi.org/10.1115/1.4036465},
    eprint = {https://asmedigitalcollection.asme.org/vibrationacoustics/article-pdf/139/5/051004/6348633/vib_139_05_051004.pdf},
}

@article{deng2024light,
  title={Light-steerable locomotion using zero-elastic-energy modes},
  author={Deng, Zixuan and Li, Kai and Priimagi, Arri and Zeng, Hao},
  journal={Nature Materials},
  volume={23},
  number={12},
  pages={1728--1735},
  year={2024},
  publisher={Nature Publishing Group UK London}
}

@article{ZHOU2015121,
title = {A torsion quasi-zero stiffness vibration isolator},
journal = {Journal of Sound and Vibration},
volume = {338},
pages = {121-133},
year = {2015},
issn = {0022-460X},
doi = {https://doi.org/10.1016/j.jsv.2014.10.027},
url = {https://www.sciencedirect.com/science/article/pii/S0022460X14008426},
author = {Jiaxi Zhou and Daolin Xu and Steven Bishop}
}

@article{LIU20133359,
title = {On the characteristics of a quasi-zero stiffness isolator using Euler buckled beam as negative stiffness corrector},
journal = {Journal of Sound and Vibration},
volume = {332},
number = {14},
pages = {3359-3376},
year = {2013},
issn = {0022-460X},
doi = {https://doi.org/10.1016/j.jsv.2012.10.037},
url = {https://www.sciencedirect.com/science/article/pii/S0022460X13000813},
author = {Xingtian Liu and Xiuchang Huang and Hongxing Hua}
}

@article{carrella2007static,
  title={Static analysis of a passive vibration isolator with quasi-zero-stiffness characteristic},
  author={Carrella, A and Brennan, MJ and Waters, TP},
  journal={Journal of sound and vibration},
  volume={301},
  number={3-5},
  pages={678--689},
  year={2007},
  publisher={Elsevier}
}

@article{mustaza2018stiffness,
  title={Stiffness control for soft surgical manipulators},
  author={Mustaza, Seri M and Saaj, Chakravarthini M and Comin, Francisco J and Albukhanajer, Wissam A and Mahdi, Duale and Lekakou, Constantina},
  journal={International Journal of Humanoid Robotics},
  volume={15},
  number={05},
  pages={1850021},
  year={2018},
  publisher={World Scientific}
}

@article{kulic2005,
doi = {10.1209/epl/i2005-10273-1},
url = {https://doi.org/10.1209/epl/i2005-10273-1},
year = {2005},
month = {oct},
publisher = {},
volume = {72},
number = {4},
pages = {527},
author = {I. M. Kulić and R. Thaokar and H. Schiessel},
title = {Twirling DNA rings—Swimming nanomotors ready for a kickstart},
journal = {Europhysics Letters}
}

@article{Schenk2014,
author = {Mark Schenk and Simon D Guest},
title ={On zero stiffness},
journal = {Proceedings of the Institution of Mechanical Engineers, Part C: Journal of Mechanical Engineering Science},
volume = {228},
number = {10},
pages = {1701-1714},
year = {2014},
doi = {10.1177/0954406213511903},
URL = {https://doi.org/10.1177/0954406213511903}
}

@article{guest2011zero,
  title={A zero-stiffness elastic shell structure},
  author={Guest, Simon and Kebadze, Elizbar and Pellegrino, Sergio},
  journal={Journal of Mechanics of Materials and Structures},
  volume={6},
  number={1},
  pages={203--212},
  year={2011},
  publisher={Mathematical Sciences Publishers}
}

@article{10.1115/1.4011755,
    author = {Yamaki, N.},
    title = {Buckling of a Thin Annular Plate Under Uniform Compression},
    journal = {Journal of Applied Mechanics},
    volume = {25},
    number = {2},
    pages = {267-273},
    year = {2021},
    month = {06},
    issn = {0021-8936},
    doi = {10.1115/1.4011755},
    url = {https://doi.org/10.1115/1.4011755},
    eprint = {https://asmedigitalcollection.asme.org/appliedmechanics/article-pdf/25/2/267/6750831/267_1.pdf},
}

@article{bedroud2015axisymmetric,
  title={Axisymmetric/asymmetric buckling of functionally graded circular/annular Mindlin nanoplates via nonlocal elasticity},
  author={Bedroud, Mohammad and Nazemnezhad, Reza and Hosseini-Hashemi, Shahrokh},
  journal={Meccanica},
  volume={50},
  number={7},
  pages={1791--1806},
  year={2015},
  publisher={Springer}
}

@article{broucke1973periodic,
  title={Periodic solutions of a spring-pendulum system},
  author={Broucke, R and Baxa, PA},
  journal={Celestial mechanics},
  volume={8},
  number={2},
  pages={261--267},
  year={1973},
  publisher={Springer}
}

@article{duka2019elastic,
  title={On the elastic pendulum, parametric resonance and ‘pumping’swings},
  author={Duka, Bejo and Duka, Raimonda},
  journal={European Journal of Physics},
  volume={40},
  number={2},
  pages={025005},
  year={2019},
  publisher={IOP Publishing}
}

@article{warminski2006autoparametric,
  title={Autoparametric vibrations of a nonlinear system with pendulum},
  author={Warminski, J and Kecik, K},
  journal={Mathematical Problems in Engineering},
  volume={2006},
  number={1},
  pages={080705},
  year={2006},
  publisher={Wiley Online Library}
}

@article{VANDERWEELE1996245,
title = {The order—chaos—order sequence in the spring pendulum},
journal = {Physica A: Statistical Mechanics and its Applications},
volume = {228},
number = {1},
pages = {245-272},
year = {1996},
issn = {0378-4371},
doi = {https://doi.org/10.1016/0378-4371(95)00426-2},
url = {https://www.sciencedirect.com/science/article/pii/0378437195004262},
author = {J.P. {van der Weele} and E. {de Kleine}}
}

@article{10.1007/s00542-016-3210-y,
author = {Golmakani, M. E. and Vahabi, H.},
title = {Nonlocal buckling analysis of functionally graded annular nanoplates in an elastic medium with various boundary conditions},
year = {2017},
issue_date = {August    2017},
publisher = {Springer-Verlag},
address = {Berlin, Heidelberg},
volume = {23},
number = {8},
issn = {0946-7076},
url = {https://doi.org/10.1007/s00542-016-3210-y},
doi = {10.1007/s00542-016-3210-y},
journal = {Microsyst. Technol.},
month = aug,
pages = {3613–3628},
numpages = {16}
}

@article{10.1098/rspa.2017.0364,
    author = {Hamouche, W. and Maurini, C. and Vidoli, S. and Vincenti, A.},
    title = {Multi-parameter actuation of a neutrally stable shell: a flexible gear-less motor},
    journal = {Proceedings of the Royal Society A: Mathematical, Physical and Engineering Sciences},
    volume = {473},
    number = {2204},
    pages = {20170364},
    year = {2017},
    month = {08},
    issn = {1364-5021},
    doi = {10.1098/rspa.2017.0364},
    url = {https://doi.org/10.1098/rspa.2017.0364},
    eprint = {https://royalsocietypublishing.org/rspa/article-pdf/doi/10.1098/rspa.2017.0364/365598/rspa.2017.0364.pdf},
}

@article{MORIMOTO2020105610,
title = {Elastic buckling of a free-standing annulus subjected to partial shrinkage},
journal = {International Journal of Mechanical Sciences},
volume = {177},
pages = {105610},
year = {2020},
issn = {0020-7403},
doi = {https://doi.org/10.1016/j.ijmecsci.2020.105610},
url = {https://www.sciencedirect.com/science/article/pii/S0020740319345333},
author = {Takuya Morimoto and Fumihiro Ashida and Takahiro Tomita}
}

@article{KOUTSOGIANNAKIS2023104745,
title = {Double restabilization and design of force–displacement response of the extensible elastica with movable constraints},
journal = {European Journal of Mechanics - A/Solids},
volume = {100},
pages = {104745},
year = {2023},
issn = {0997-7538},
doi = {https://doi.org/10.1016/j.euromechsol.2022.104745},
url = {https://www.sciencedirect.com/science/article/pii/S0997753822001887},
author = {P. Koutsogiannakis and D. Bigoni and F. {Dal Corso}}
}

@article{turco2018discrete,
  title={Discrete is it enough? The revival of Piola--Hencky keynotes to analyze three-dimensional Elastica},
  author={Turco, Emilio},
  journal={Continuum Mechanics and Thermodynamics},
  volume={30},
  number={5},
  pages={1039--1057},
  year={2018},
  publisher={Springer}
}

@article{turco2020lagrangian,
  title={A Lagrangian Hencky-type non-linear model suitable for metamaterials design of shearable and extensible slender deformable bodies alternative to Timoshenko theory},
  author={Turco, Emilio and Barchiesi, Emilio and Giorgio, Ivan and Dell’Isola, Francesco},
  journal={International Journal of Non-Linear Mechanics},
  volume={123},
  pages={103481},
  year={2020},
  publisher={Elsevier}
}

\appendix

\section{Nomenclature}
\label{app:nomenclature}

\setcounter{equation}{0}
\renewcommand{\theequation}{A.\arabic{equation}}

For completeness, the main dimensional and nondimensional quantities used
throughout the manuscript are summarized in Tables~\ref{tab:dimensional}
and~\ref{tab:nondimensional}, respectively. The extension stiffness of the
foundation, $k^{(+)}$, is used as the reference stiffness in the
nondimensionalization. The compression-to-extension stiffness ratio is
defined as $\gamma=k^{(-)}/k^{(+)}$.

\begin{table}[H]
    \centering
    \caption{Dimensional quantities used in the formulation.}
    \label{tab:dimensional}
    \begin{tabular}{c p{0.58\textwidth} c}
        \hline
        Symbol & Physical meaning & Units \\
        \hline
        $m$
        & Mass of each discrete element
        & $\mathrm{kg}$ \\

        $\ell_0$
        & Natural length of the circumferential linear springs
        & $\mathrm{m}$ \\

        $\ell$
        & Distance between neighboring masses in the compressed
          undeformed configuration, $\ell=\ell_0(1+p)$
        & $\mathrm{m}$ \\

        $k_c$
        & Stiffness of the circumferential linear springs
        & $\mathrm{N\,m^{-1}}$ \\

        $k_r$
        & Stiffness of the torsional springs
        & $\mathrm{N\,m}$ \\

        $k^{(+)}$
        & Foundation stiffness when the corresponding spring is in extension;
          used as the reference stiffness in the nondimensionalization
        & $\mathrm{N\,m^{-1}}$ \\

        $k^{(-)}$
        & Foundation stiffness when the corresponding spring is in compression
        & $\mathrm{N\,m^{-1}}$ \\

        $c$
        & Linear viscous damping coefficient
        & $\mathrm{N\,s\,m^{-1}}$ \\

        $u_i$
        & Vertical displacement of the $i$th mass
        & $\mathrm{m}$ \\

        $u_0(t)$
        & Prescribed vertical displacement of the foundation,
          $u_0(t)=\hat{v}_0\ell_0\sin(\omega t)$
        & $\mathrm{m}$ \\

        $t$
        & Dimensional time
        & $\mathrm{s}$ \\

        $\omega$
        & Angular frequency of the harmonic foundation excitation
        & $\mathrm{rad\,s^{-1}}$ \\

        \hline
    \end{tabular}
\end{table}

\begin{longtable}{c p{0.28\textwidth} p{0.44\textwidth}}
    \caption{Nondimensional parameters and kinematic quantities used in the formulation.}
    \label{tab:nondimensional}\\
    \hline
    Symbol & Definition & Physical meaning \\
    \hline
    \endfirsthead

    \multicolumn{3}{c}%
    {{\tablename\ \thetable{} -- continued from previous page}}\\
    \hline
    Symbol & Definition & Physical meaning \\
    \hline
    \endhead

    \hline
    \multicolumn{3}{r}{{Continued on next page}}\\
    \endfoot

    \hline
    \endlastfoot

    $N$
    & ---
    & Number of discrete masses in the ring \\

    $n$
    & ---
    & Buckling mode number \\
    
    $\alpha$
    & $\displaystyle \alpha=\frac{2\pi}{N}$
    & Angular spacing between two neighboring masses \\
    
    $\tau$
    & $\displaystyle \tau=\omega t$
    & Nondimensional time \\
    
    $p$
    & $\displaystyle p=\frac{\ell}{\ell_0}-1$
    & Compression parameter; $p=0$ corresponds to the uncompressed
      configuration and increasingly negative values correspond to
      increasing compression \\
      
    $\rho$
    & $\displaystyle \rho=\frac{m\omega^2}{k^{(+)}}$
    & Nondimensional inertial parameter \\

    $q$
    & $\displaystyle q=\frac{k_r}{k^{(+)}\ell_0^2}$
    & Nondimensional torsional stiffness \\

    $w$
    & $\displaystyle w=\frac{k_c}{k^{(+)}}$
    & Ratio between circumferential and foundation stiffnesses \\

    $\zeta$
    & $\displaystyle \zeta=\frac{c\omega}{k^{(+)}}$
    & Nondimensional damping parameter \\
    
    $v_i$
    & $\displaystyle v_i=\frac{u_i}{\ell_0}$
    & Nondimensional vertical displacement of the $i$-th mass \\

    $v_0$
    & $\displaystyle v_0=\frac{u_0}{\ell_0}
      =\hat{v}_0\sin\tau$
    & Nondimensional displacement of the foundation \\

    $\hat{v}_0$
    & ---
    & Nondimensional amplitude of the foundation motion; the corresponding
      dimensional displacement amplitude is $\hat{v}_0\ell_0$ \\

    $\lambda_i$
    &
    $\displaystyle
    \lambda_i=
    \sqrt{(1+p)^2+(v_i-v_{i-1})^2}$
    & Nondimensional distance between neighboring masses \\

    $\phi_i$
    &
    $\displaystyle
    \phi_i=
    \tan^{-1}
    \left(
    \frac{v_i-v_{i-1}}{1+p}
    \right)$
    & Inclination angle of the $i$th circumferential spring \\ \vspace{0.5em} \\

    $\gamma$
    & $\displaystyle \gamma=\frac{k^{(-)}}{k^{(+)}}$
    & Compression-to-extension foundation stiffness ratio\\ \vspace{0em}\\

    $\kappa_i$
    &
    $\displaystyle
    \kappa_i=
    \begin{cases}
        1,      & v_i-v_0\geq 0,\\
        \gamma, & v_i-v_0<0
    \end{cases}$
    & Instantaneous nondimensional stiffness of the $i$th foundation
      spring \\ \\

    $z$
    & $\displaystyle z=\frac{1}{N}\sum_{i=1}^{N}v_i$
    & Nondimensional rigid-body vertical translation of the ring \\ \\

    $A$
    &
    $\displaystyle
    A=\frac{\max_i(v_i)-\min_i(v_i)}{2}$
    & Nondimensional amplitude of the buckled deformation \\

    $\theta$
    & ---
    & Phase coordinate describing the rotation of the buckled
      deformation; the physical orientation of the deformation changes
      by $\theta/n$ \\

    $p_{\mathrm{cr}}$
    & ---
    & Critical value of the compression parameter at buckling \\

    $A_{\mathrm{eq}}$
    & ---
    & Equilibrium amplitude of the buckled configuration \\

    $z_{\mathrm{eq}}$
    & ---
    & Equilibrium vertical translation of the ring \\

    $\theta_{\mathrm{eq}}$
    & ---
    & Equilibrium rotational phase of the buckled configuration \\

    $\accentset{\ast}{(\cdot)}$
    & $\displaystyle \accentset{\ast}{(\cdot)}=\frac{d(\cdot)}{d\tau}$
    & Differentiation with respect to nondimensional time \\  \\

    $\accentset{\ast\ast}{(\cdot)}$
    & $\displaystyle \accentset{\ast\ast}{(\cdot)}=\frac{d^2(\cdot)}{d\tau^2}$
    & Second differentiation with respect to nondimensional time \\

\end{longtable}

\section{Reduced-order kinematic model}\label{app:red_kin_derivation}

\setcounter{equation}{0}
\renewcommand{\theequation}{B.\arabic{equation}}

The equations of motion under the constrained kinematic model given by Eq. \ref{eq:eig_sol} can be obtained by projecting the unconstrained dynamic model on the kinematic model. To do this, we first write the system of equations in a weak form and then we substitute our kinematic model $v_i=v_i(z,A,\theta)$.

By performing the variation of the Lagrangian $\mathcal L=\mathcal T-\mathcal V$, with $\mathcal T$, and $\mathcal V$ defined in Eq. \ref{energies}. The weak form of the problem is
\begin{equation}\label{weak}
\begin{aligned}
    \sum_{i=1}^N \left\{ (\rho\accentset{\ast\ast}{v}_i+v_i-v_0)\, \delta v_i  + w\left( 1 - \frac{1}{\lambda_i} \right) \right.
    & (v_i-v_{i-1})(\delta v_i-\delta v_{i-1})\\ 
    & \left. +\ q (1+p) (\phi_{i+1}-\phi_i) \left( \frac{\delta v_{i+1}-\delta v_i}{\lambda_{i+1}^2} - \frac{\delta v_{i}-\delta v_{i-1}}{\lambda_{i}^2} \right) \right\}=0.
\end{aligned}
\end{equation}
Under the constrained kinematics framework, the admissible variations of $v_i$ take the form
\begin{equation}
    \delta v_i = \delta z + \cos(a n i +\theta)\, \delta A - A \sin(a n i +\theta)\delta \theta.
\end{equation}
By expanding the trigonometric functions and defining the $1\times 2$ matrix
\begin{equation}
    Y_i = [ \cos(a n i), -\sin(a n i) ],
\end{equation}
and defining the derivative of $v_i$ with respect to $\{z,A,\theta\}$
\begin{equation}
    \mbox{D}v_i = \begin{pmatrix}\displaystyle \frac{\partial v_i}{\partial z} \\\displaystyle\frac{\partial v_i}{\partial A} \\\displaystyle \frac{\partial v_i}{\partial \theta} \end{pmatrix}
    = \begin{pmatrix} 1 \\ \mathbf Y_i \mathbf R(\theta) (1,0)^T \\ \mathbf Y_i \mathbf R(\theta) (0,1)^T A \end{pmatrix}
\end{equation}
we can write $v_i$, its time derivatives, and its variation as
\begin{equation}
\begin{aligned}
    & v_i = z + Y_i \cdot R(\theta)\cdot\begin{pmatrix}A\\0\end{pmatrix} \\
    & \accentset{\ast}{v}_i = \accentset{\ast}z + Y_i \cdot R(\theta)\cdot\begin{pmatrix}\accentset{\ast}A\\A\accentset{\ast}\theta\end{pmatrix}\\
    & \accentset{\ast\ast}{v}_i = \accentset{\ast\ast}z + Y_i \cdot R(\theta)\cdot\begin{pmatrix}\accentset{\ast\ast}A-A\accentset{\ast}\theta^2\\A\accentset{\ast\ast}\theta+2\accentset{\ast}A\accentset{\ast}\theta\end{pmatrix}\\
    & \delta v_i = \left\{1, Y_i\cdot R(\theta)\begin{pmatrix}1\\0\end{pmatrix},  Y_i\cdot R(\theta)\begin{pmatrix}0\\A\end{pmatrix} \right\}\cdot \begin{pmatrix}\delta z\\\delta A\\\delta \theta\end{pmatrix}
\end{aligned}
\end{equation}
where $\mathbf R(\theta)$ is the $2\times 2$ rotation matrix.
Further, the difference between these quantities can be calculated as
\begin{equation}
\begin{aligned}
    & v_i-v_{i-1} = Y_i \cdot R(\theta) \begin{pmatrix}2 A \sin^2(an/2)\\A\sin(a n)\end{pmatrix} \\
    & \delta v_i-\delta v_{i-1} = \left\{ 0, Y_i \cdot R(\theta) \begin{pmatrix}1-\cos(a n)\\\sin(a n)\end{pmatrix},  Y_i \cdot R(\theta) \begin{pmatrix}-\sin(a n)\\1-\cos(a n)\end{pmatrix} A \right\}\cdot \begin{pmatrix}\delta z\\\delta A\\\delta \theta\end{pmatrix} \\
    & \delta v_{i+1}-\delta v_{i} = \left\{ 0, Y_i \cdot R(\theta) \begin{pmatrix}\cos(a n)-1\\\sin(a n)\end{pmatrix},  Y_i \cdot R(\theta) \begin{pmatrix}-\sin(a n)\\\cos(a n)-1\end{pmatrix} A \right\}\cdot \begin{pmatrix}\delta z\\\delta A\\\delta \theta\end{pmatrix}.
\end{aligned}
\end{equation}
By substituting these quantities in Eq. \ref{weak} and requiring it to be valid for all variations $\delta z$, $\delta A$, $\delta\theta$, we obtain the following system of equations
\begin{equation}\label{eom_con_mat}
    \mathbf M_0 \accentset{\ast\ast}z + \mathbf K_0 (z-v_0) + \mathbf M_1 \mathbf R(\theta) \begin{pmatrix}\accentset{\ast\ast}A-A\accentset{\ast}\theta^2\\A\accentset{\ast\ast}\theta+2\accentset{\ast}A\accentset{\ast}\theta\end{pmatrix} +
    \mathbf K_1 \mathbf R(\theta) \begin{pmatrix}1\\0\end{pmatrix}A + \mathbf K_2 \mathbf R(\theta) \begin{pmatrix}2\sin^2\frac{a n}{2}\\\sin a n\end{pmatrix}A + \mathbf G = 0,
\end{equation}
where
\begin{equation}
    \begin{aligned}
        & \mathbf M_0 = \rho \sum_{i=1}^N \mbox{D}v_i, \quad \mathbf M_1 = \rho \sum_{i=1}^N \mbox{D}v_i \mathbf Y_i, \\
        & \mathbf K_0 = \sum_{i=1}^N \mbox{D}v_i, \quad \mathbf K_1 = \sum_{i=1}^N \mbox{D}v_i \mathbf Y_i, \quad \mathbf K_2 = w \sum_{i=1}^N (\mbox{D}v_i-\mbox{D}v_{i-1}) \mathbf Y_i, \\
        & \begin{aligned}\mathbf B(A,\theta) = &-w \left\{\sum_{i=1}^N (\mbox Dv_i-\mbox Dv_{i-1})\mathbf Y_i \frac{1}{\lambda_i}  \right\} \mathbf R(\theta) \begin{pmatrix}
            2\sin^2\frac{a n}{2} \\ \sin a n
        \end{pmatrix} A \\
        &+ q (1+p) \left\{ \sum_{i=1}^N (\mbox Dv_{i+1}-\mbox Dv_{i}) \frac{\phi_{i+1}-\phi_i}{\lambda_{i+1}^2} \right\} - q (1+p) \left\{ \sum_{i=1}^N (\mbox Dv_{i}-\mbox Dv_{i-1}) \frac{\phi_{i+1}-\phi_i}{\lambda_{i}^2} \right\}.
        \end{aligned}
    \end{aligned}
\end{equation}

We notice the following properties
\begin{equation}
    \sum_{i=1}^N \mathbf Y_i = \mathbf 0, \quad 
    \sum_{i=1}^N \mathbf Y_i^T \mathbf Y_i = \mbox{diag}\left( \sum_{i=1}^N \cos^2 (an i),\ \sum_{i=1}^N \sin^2 (an i) \right),
\end{equation}
\begin{equation}
    \frac{\partial v_i}{\partial z}-\frac{\partial v_{i-1}}{\partial z} = \frac{\partial v_{i+1}}{\partial z}-\frac{\partial v_{i}}{\partial z} = 0.
\end{equation}
Using these properties, and simplifying the summations where possible, the first of the three equations \ref{eom_con_mat} becomes
\begin{equation}\label{eom_con_1}
    \rho \accentset{\ast\ast}z + z-v_0 = 0.
\end{equation}
Subsequently, we can treat the other two equations together and after some algebraic manipulation they can be written as
\begin{equation}
    \mathbf R(\theta) \mathbf R(\theta) \begin{pmatrix}\accentset{\ast\ast}A-A\accentset{\ast}\theta^2\\A\accentset{\ast\ast}\theta+2\accentset{\ast}A\accentset{\ast}\theta\end{pmatrix} + \frac{1}{\rho} \mathbf R(\theta) \mathbf R(\theta)\begin{pmatrix}A\\0\end{pmatrix} + \frac{1}{\rho} \begin{bmatrix}
        2\sin^2\frac{a n}{2} & -\sin a n \\ \sin a n & 2 \sin^2\frac{a n}{2}
    \end{bmatrix} \mathbf R(\theta) \mathbf R(\theta) \begin{pmatrix}
        2 \sin^2\frac{a n}{2} \\ \sin a n
    \end{pmatrix} A + \frac{2}{\rho N}\begin{pmatrix}
        B_2 \\ B_3/A
    \end{pmatrix} = \mathbf 0.
\end{equation}
Finally, multiplying with $(\mathbf R(\theta) \mathbf R(\theta))^{-1}$ we get the following two equations
\begin{equation}\label{eom_con_2}
    \begin{aligned}
        & \accentset{\ast\ast}A + \left(\frac{2-2\cos a n +\cos(2 a n)}{\rho}-\accentset{\ast}\theta^2\right) A + G_2(A,\theta) = 0 \\
        & A\accentset{\ast\ast}\theta+2\accentset{\ast}A\accentset{\ast}\theta + 4 \sin^2\frac{an}{2} \sin(a n ) A + G_3(A,\theta) = 0,
    \end{aligned}
\end{equation}
with
\begin{equation}
    \begin{pmatrix}
        G_2(A,\theta) \\ G_3(A,\theta)
    \end{pmatrix} = \frac{2}{\rho N} (\mathbf R(\theta) \mathbf R(\theta))^{-1} \begin{pmatrix}
        B_2(A,\theta) \\ B_3(A,\theta) / A
    \end{pmatrix}.
\end{equation}
It can be observed that Eq. \ref{eom_con_1} is independent of $A$, and $\theta$, while Eqs. \ref{eom_con_2} are independent of $z$, thus making the first decoupled from the latter.

\section{Numerical integration scheme} \label{app:numerical}

\setcounter{equation}{0}
\renewcommand{\theequation}{C.\arabic{equation}}

The equations of motion given in Eq. \ref{eq:eom_nd} are integrated in time using a numerical scheme based on the Newmark method. In more detail, starting from $\tau_0=0$, the quantities $v_i$, $\accentset{\ast}v_i$, and $\accentset{\ast\ast}v_i$ are calculated for times $\tau_k=\tau_0+k\, \Delta\tau$. The system of equations is written for a given timeframe $k$ as
\begin{equation}
    \mathbf f\!\left(\mathbf v, \accentset{\ast}{\mathbf v}, \accentset{\ast\ast}{\mathbf v}\right) = 0,
\end{equation}
where
\begin{equation}
    \mathbf v=\{v_1,v_2,\dots,v_N\}^T,\quad \accentset{\ast}{\mathbf v}=\{\accentset{\ast}v_1,\accentset{\ast}v_2,\dots,\accentset{\ast}v_N\}^T
    ,\quad \accentset{\ast\ast}{\mathbf v}=\{\accentset{\ast\ast}v_1,\accentset{\ast\ast}v_2,\dots,\accentset{\ast\ast}v_N\}^T.
\end{equation}
Then, by approximating $\accentset{\ast}{\mathbf v}_k$, and $\mathbf v_k$ as
\begin{equation}\label{newmark_assumption}
\begin{aligned}
    & \accentset{\ast}{\mathbf v}_k = \accentset{\ast}{\mathbf v}_{k-1} + (1-\beta_2) \Delta\tau \accentset{\ast\ast}{\mathbf v}_{k-1} + \beta_2 \Delta\tau \accentset{\ast\ast}{\mathbf v}_{k}, \\
    & {\mathbf v}_k = {\mathbf v}_{k-1} + \Delta\tau \accentset{\ast}{\mathbf v}_{k-1} + \frac{(\Delta\tau)^2}{2}\left[ (1-2\beta_1) \accentset{\ast\ast}{\mathbf v}_{k-1} + 2\beta_1 \accentset{\ast\ast}{\mathbf v}_{k} \right],
\end{aligned}
\end{equation}
we can write the equations of motion as a function of $\accentset{\ast\ast}{\mathbf v}_{k}$ only, assuming all three quantities are known for the time instant $\tau_{k-1}$. After solving for $\accentset{\ast\ast}{\mathbf v}_{k}$, quantites ${\mathbf v}_k$, and $\accentset{\ast}{\mathbf v}_k$ can be calculated from Eq. \ref{newmark_assumption}. For the purposes of obtaining the results shown in the main text, the numerical integration parameters used were $\beta_1=0.25$, $\beta_2=0.5$, $\Delta\tau=\frac{2\pi}{40}$. This set of parameters ensures energy conservation and unconditional stability.

The Newmark method for $\beta_1=0.25$, $\beta_2=0.5$ is second-order accurate. The temporal discretization error is $e=O(\Delta\tau^2)$. The chosen timestep size $\Delta\tau=\frac{2\pi}{40}$ yields an error in the order $e=O((2\pi/40)^2)\approx O(0.02)$. This choice offers a good balance between accuracy and computation speed. 



\end{document}